\documentstyle[emulateapj,apjfonts]{article}

\received{}
\accepted{}

\begin{document}

\submitted{Accepted for ApJ, 22 January 1999}

\title{The twisted parsec-scale structure of 0735+178}

\author{Jos\'e-Luis G\'omez}
\affil{Instituto de Astrof\'{\i}sica de Andaluc\'{\i}a, CSIC,
Apartado 3004, 18080 Granada, Spain}
\authoremail{jlgomez@iaa.es}

\author{Alan P. Marscher}
\affil{Department of Astronomy, Boston University, 725 Commonwealth
Avenue, Boston, MA 02215, USA}
\authoremail{marscher@buast0.bu.edu}

\author{Antonio Alberdi}
\affil{Instituto de Astrof\'{\i}sica de Andaluc\'{\i}a, CSIC,
Apartado 3004, 18080 Granada, Spain}
\authoremail{antxon@iaa.es}

\author{Denise C. Gabuzda}
\affil{Astro Space Center, P. N. Lebedev Physical Institute,
Leninsky Prospekt 53, 117924, Moscow, Russia}
\authoremail{gabuzda@sci.lebedev.ru}

\begin{abstract}

  We present two epoch polarimetric images of the BL~Lac object 0735+178
obtained with the Very Long Baseline Array at 22 and 43 GHz. These images
provide the highest resolution observations of this source to date, and show a
twisted jet with two sharp apparent bends of 90\( ^{\circ } \) within two
milliarcseconds of the core. The magnetic field appears to smoothly follow one
of the bends in the jet, suggesting that this structure may be the result of a
precessing nozzle in the jet of 0735+178. Proper motion information is not
conclusive as to whether the twisted geometry is the result of knots moving
ballistically or following the direction of the jet axis.

\end{abstract}


\section{Introduction}

  0735+178 was first classified as a BL~Lacertae object by Carswell et
al. (\cite{carswell 74}), who identified a pair of absorption lines in an
otherwise featureless optical spectrum, thereby obtaining a lower limit to the
redshift of the source $z_{\rm min}=0.424$. Hence, any observed apparent
speeds in 0735+178 calculated using this redshift are strictly lower limits.

  Radio maps of this source at milliarcsecond resolution have been obtained by
using VLBI arrays at centimeter wavelengths, and show a compact core and a jet
of emission extending to the northeast. Polarimetric VLBI observations of this
source were first performed by Gabuzda, Wardle \& Roberts (\cite{De89}) at a
wavelength of 6 cm: the corresponding polarized intensity image shows a jet
magnetic field that is predominantly perpendicular to the jet axis, together
with a dramatic change in the polarized flux by 40\% over a period of 24
hours.  The multi-epoch 6 cm VLBI observations of 0735+178 presented by
B\"a\"ath \& Zhang (\cite{BZ91}) indicated superluminal motion in their
component \emph{C0} at an apparent velocity of \( \sim \)7.9 \( h^{-1}c \) (\(
H_{\circ } \)= 100 \( h \) km s\( ^{-1} \)Mpc\( ^{-1} \), \( q_{\circ } \)=
0.5), moving between the core and a stationary component, which they labeled
\emph{B}, situated about 4.2 mas from the core. This situation resembles that
found in the quasar 4C~39.25 (Alberdi et al. \cite{An93} and references
therein), where a very strong component moves superluminally between the core
and an outer stationary component, and its increasing brightness as it moves
is interpreted as Doppler enhancement caused by a bend toward the line of
sight. Gabuzda et al. (\cite{De94}; hereafter G94) confirmed the motion of the
superluminal component \emph{C0}, which they designated as \emph{K2}, at a
velocity of 7.4 \( h^{-1}\, c \), and detected two new superluminal components
with apparent transverse speeds of 5.0 \( h^{-1}\, c \) and 4.2 \( h^{-1}\, c
\). The observations of G94 were near the time of intersection of the moving
component \emph{K2} with the stationary one \emph{K1} (component \emph{B} of
B\"a\"ath \& Zhang \cite{BZ91}), which took place at epoch \( \sim
\)1989.8. Their images do not show any evidence for a violent interaction
between these components, and they interpret this in terms of a model similar
to that suggested by B\"a\"ath \& Zhang (\cite{BZ91}), in which the stationary
component \emph{K1} is associated with a bend in the jet toward the line of
sight. However, if this is the case, we might expect a deceleration of
\emph{K2} as it approaches \emph{K1}, along with an increase of its flux due
to enhancement of Doppler boosting, as observed and simulated in the case of
4C~39.25 (Alberdi et al. \cite{An93}). Neither of these were observed in
0735+178. If \emph{K1} corresponds to a bend toward the observer, it also
remains unclear why G94 measured component \emph{K1} to have shifted \( \sim
\) 0.7 mas outward sometime between 1981.9 and 1983.5, after its collision
with \emph{K2}, and to have remained in that position afterward. One
explanation for this ``dragging'' of \emph{K1} is that we are dealing with an
interaction between a moving shock (\emph{K2}) and a standing conical shock
(\emph{K1}), as G\'omez et al. (\cite{Go97}) have modeled through numerical
simulations. In this scenario, an increase in the Mach number of the flow
causes the stationary shock to reset to a new site downstream of its original
position.

  Higher resolution VLBI maps obtained by B\"a\"ath et al. (\cite{Bal91}) at
1.3 cm reveal a complex structure consisting of several distinct superluminal
components moving in a rather straight jet extending to the
northeast. However, a fit to the kinematical properties of the superluminal
components led B\"a\"ath et al. (\cite{Bal91}) to suggest a bend of the jet in
the inner milliarcsecond.  Indeed, a quite complex structure in the inner
milliarcsecond of 0735+178 appears in the 1.3 cm VLBI maps presented by Zhang
\& B\"a\"ath (\cite{ZB91}). Over a time span of about 1.2 yrs, the jet
structural position angle was found to change from \( \sim \) 25\( ^{\circ }
\) to \( \sim \) 73\( ^{\circ } \).  It is unclear whether Zhang \& B\"a\"ath
(\cite{ZB91}) observed different components tracing a bent common path, or
components being ejected from the core with different position angles, as G94
suggested based on their observations. The first direct evidence of a curved
structure in the inner jet of 0735+178 was presented by Kellermann et
al. (\cite{Ke98}), as part of a Very Long Baseline Array survey
(VLBA\footnote{ The VLBA is an instrument of the National Radio Astronomy
Observatory, which is a facility of the National Science Foundation operated
under cooperative agreement by Associated Universities, Inc.  }) at 15 GHz.

  In this paper we present the first polarimetric VLBI observations of
0735+178 at 1.3 cm and 7 mm, obtained with the VLBA. The images show an
apparently quite twisted structure in the inner two milliarcseconds of the
jet, in very good agreement with the total intensity 2 cm VLBA image obtained
by Kellermann et al. (\cite{Ke98}). We analyze and interpret the magnetic
field struture and jet morphology, and discuss the possibility of a precessing
jet in 0735+178.

\section{Observations}

  The observations were performed on 1996 November 11 and December 22 using
the VLBA at 1.3 cm and 7 mm in snapshot mode. The data were recorded in 1-bit
sampling VLBA format with 32~MHz bandwidth per circular polarization. The
reduction of the data was performed within the NRAO Astronomical Image
Processing System (AIPS) software in the usual manner (e.g., Lepp\"anen et
al. \cite{Le95}).  Delay differences between the right- and left-handed
systems were estimated over a short scan of cross-polarized data of a strong
calibrator (3C~454.3).  The instrumental polarization was determined using the
feed solution algorithm developed by Lepp\"anen et al. (\cite{Le95}). We refer
the readers to G\'omez et al. (\cite{Go98}) for further details about the
reduction and calibration of the data.

  Due to poor weather conditions in some of the antenna locations during the
second run, specially at Brewster, much of the 7 mm data had to be deleted,
leading to a significantly poorer total intensity image and the loss of all
polarization at this epoch.

\section{Results}

\begin{figure*}
\plotone{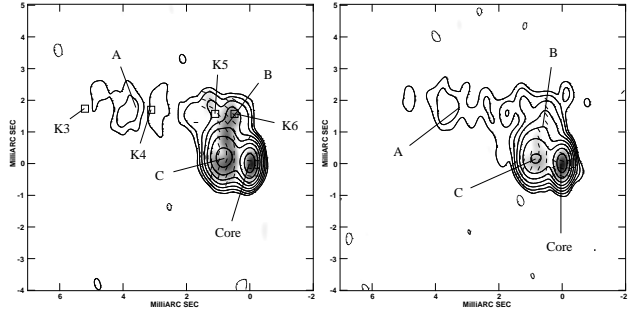}
\figcaption{VLBA images of 0735+178 at 1.3 cm at epochs 1996 November 11
(\emph{left}) and 1996 December 22 (\emph{right}). Total intensity is plotted
as contour maps, while the linear gray scale shows the linearly polarized
intensity. The superposed sticks give the direction of the \emph{magnetic
field} vector. Contour levels are -0.5, 0.5, 1, 2, 4, 8, 16, 32, and 64\% of
the peaks of 335 (November 1996) and 322 (December 1996) mJy beam\protect\(
^{-1}\protect \). For November 1996 the convolving beam is 0.74 \protect\(
\times \protect \) 0.38 mas (FWHM), with position angle -1.5\protect\( ^{\circ
}\protect \), and the maximum in polarized intensity is 7.6 mJy beam\protect\(
^{-1}\protect \), with a noise level of 1.5 mJy beam\protect\( ^{-1}\protect
\). For December 1996 the convolving beam is 0.73 \protect\( \times \protect
\) 0.36 mas (FWHM), with position angle -2.3\protect\( ^{\circ }\protect \),
and the maximum in polarized intensity is 12.9 mJy beam\protect\(
^{-1}\protect \), with a noise level of 2.6 mJy beam\protect\( ^{-1}\protect
\).\label{figure1}}
\end{figure*}

\begin{figure*}
\plotone{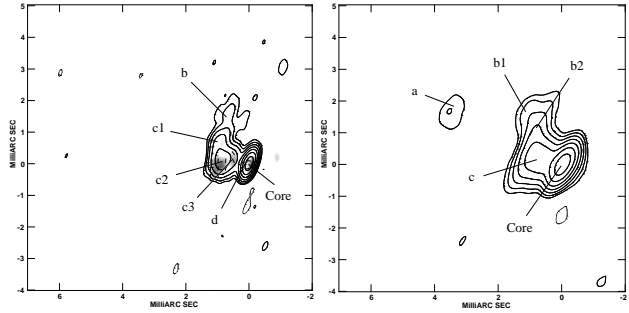}
\figcaption{Same as Fig. \ref{figure1}, but at 7 mm. Contour levels are -1, 1,
2, 4, 8, 16, 32, and 64\% of the peaks of 307 (November 1996) and 362
(December 1996) mJy beam\protect\( ^{-1}\protect \). For November 1996 the
convolving beam is 0.57 \protect\( \times \protect \) 0.24 mas (FWHM), with
position angle -20\protect\( ^{\circ }\protect \), and the maximum in
polarized intensity is 8.8 mJy beam\protect\( ^{-1}\protect \), with a noise
level of 3.5 mJy beam\protect\( ^{-1}\protect \).  For December 1996 the
convolving beam is 0.93 \protect\( \times \protect \) 0.43 mas (FWHM), with
position angle -30.6\protect\( ^{\circ }\protect \); no polarization
information was obtained.\label{figure2}}
\end{figure*}  

  Figures \ref{figure1} and \ref{figure2} show the VLBA total and linearly
polarized intensity images of 0735+178 at 1.3 cm and 7 mm,
respectively. Tables \ref{table1} and \ref{table2} summarize the physical
parameters obtained for 0735+178 at 1.3 cm and 7 mm, respectively. Tabulated
data corresponds to position in right ascension and declination relative to
the core component, total flux, polarized flux, magnetic field position angle,
degree of polarization, and separation and structural position angle relative
to the core component. Components in the total intensity maps were analyzed
by model-fitting the uv data with circular Gaussian components using the
Difmap software (Shepherd \cite{sh97}). Components are labeled from east to
west using upper-case letters for the images at 1.3 cm, and are marked in
Figs. \ref{figure1} and \ref{figure2}. Due to the potentially large
uncertainties associated with the process of model-fitting, only components
well above the noise level were considered. For example, there is some weak
evidence for the existence of components \emph{d} and \emph{c1} in the 1.3 cm
maps, but the errors in the model-fitting are sufficiently large that we have
decided not to include these in the final model. Component \emph{B} is best
fit at the second 7 mm epoch by two separate components labeled \emph{b2} and
\emph{b1}, the combined flux of which compares quite well with that of
component \emph{b} in the 7 mm image at the first epoch.  To reduce the errors
in the model-fitting we have fit the emission from the section of the jet
beyond component \emph{B} as a single component, labeled \emph{A}.

\subsection{The Twisted Structure of the Inner Jet of 0735+178}

  B\"a\"ath et al. (\cite{Bal91}) and B\"a\"ath \& Zhang (\cite{BZ91})
suggested the existence of bent structure in the jet of 0735+178 in order to
explain the apparent acceleration of superluminal components as they moved
further from the core. G94 found that components seem to move along linear
trajectories, but to be ejected with different position angles, which could
lead to apparent curved structure in the VLBI jet.

\begin{deluxetable}{lcccccccc}
\tablecolumns{9}
\tablewidth{0pc}
\tablecaption{22 GHz models for 0735+178.\label{table1}}
\tablehead{
& 
{\scriptsize \( \alpha  \) }&
{\scriptsize \( \delta  \)}&
\( I \)&
\( p \)&
\( \chi _{m} \)&
\( m \)&
\( r \)&
\( \theta  \)\nl
Component&
(mas)&
(mas)&
(mJy)&
(mJy)&
(\( ^{\circ } \))&
(\%)&
(mas)&
(\( ^{\circ } \))
}
\startdata
\multicolumn{9}{c}{1996.86}\nl
\hline
Core\dotfill&
...&
...&
354& 
7&
86&
2&
...&
...\nl
C\dotfill&
0.81&
0.19&
361&
12&
-12&
3&
0.83&
77\nl
B\dotfill&
0.7&
1.3&
87&
9&
31&
10&
1.5&
29\nl
A\dotfill&
3.6&
1.8&
37&
...&
...&
...&
4.1&
63\nl
\hline 
\multicolumn{9}{c}{1996.98 } \nl
\hline 
Core\dotfill&
...&
...&
340&
13&
72&
4&
...&
...\nl
C\dotfill&
0.82&
0.18&
304&
9&
-21&
3&
0.84&
78\nl
B\dotfill&
0.6&
1.2&
94&
3&
36&
3&
1.3&
29\nl
A\dotfill&
3.2&
1.8&
39&
...&
...&
...&
3.7&
61\nl
\enddata
\end{deluxetable}

  Our observations directly confirm the existence of a very twisted structure
in the inner region of the jet in 0735+178, in good agreement with the
observations presented by Kellermann et al. (\cite{Ke98}). In our 1.3 cm maps
we observe two sharp 90\( ^{\circ } \) bends in the projected trajectory of
the jet, one near the position of component \emph{C} and the other close to
\emph{B}. At 7 mm the bend near the region of component \emph{c2} (component
\emph{C} at 1.3 cm) is also quite evident, while the second bend can only be
traced at the second epoch, since the emission beyond component \emph{b} is
resolved out in November 1996.

  Through a comparison with our maps, we can offer an explanation for the
otherwise puzzling structure Zhang \& B\"a\"ath (\cite{ZB91}) observed in
their 1.3 cm VLBI maps. It is possible that they observed distinct components
at different positions of a common curved path, consistent with that traced in
our images.  However in the 1.3 cm VLBI map by B\"a\"ath et al. (\cite{Bal91})
they observed in the inner milliarcsecond a straight jet extending northeast
at a position angle of \( \sim \)45\( ^{\circ } \), in contrast with the
structure we observe in our images. Furthermore, G94 measured different
ejection position angles for components in 0735+178.

  In order to allow an easier comparison with our images, we have marked in
our November 1996 map of Fig. \ref{figure1} the expected position for
components \emph{K3}, \emph{K4}, \emph{K5}, and \emph{K6} of G94 using their
measured proper motions and assuming ballistic trajectories from the core
since their last observing epoch in 1992.44. For component \emph{K}5 we have
used the position reported for their 1990.47 epoch, otherwise the
extrapolation would place this component far too east of \emph{C}, outside of
the structure we have mapped. No proper motion data is available for component
\emph{K6,} but assuming a similar value to that observed for the other
components, we have used 0.3 mas yr\( ^{-1} \).  Figure \ref{figure1} shows
that this extrapolation places the G94 components within the structure found
on our images, suggesting ballistic motion with systematically changing
component ejection directions as the most plausible explanation for the curved
structure of 0735+178.

  However, the data is also consistent with motion of components following a
common curved path. Component \emph{K3} was observed to change drastically its
position angle relative to the core between 1987.41 and 1990.47, from 75\(
^{\circ } \) to 44\( ^{\circ } \) (see G94). Component \emph{K5} also
experienced a change in its position angle between 1990.47 and 1992.44, with
its velocity vector becoming more aligned toward the direction of component
\emph{C}, between the position of components \emph{c3} and \emph{c2}. The
positions of components G94 in the inner milliarcsecond are consistent with
the twisted structure we have mapped, therefore we cannot rule out the
possibility of a common curved funnel through which components flow. Note also
that this curved funnel could change with time if it were produced by
precession of the nozzle in 0735+178.  In the case of ballistic motion, the
direction of ejection of new components should have changed progressively from
54\( ^{\circ } \), to 35\( ^{\circ } \), and then to 15\( ^{\circ } \) from
1984 to 1991.25 to explain the different structural position angles observed
for the birth of \emph{K4}, \emph{K5}, and \emph{K6}. The direction of
ejection would then have needed to change to 78\( ^{\circ } \) within about 4
yr, if we assume that component \emph{C} was ejected around the beginning of
1995, coincident with an outburst in total flux at 22 and 37 GHz measured at
Mets\"ahovi (Ter\"asranta et al. \cite{Hi98}). Since then, it would have had
to remain at a similar orientation (at least on the plane of the sky) to give
birth to components \emph{c3} and \emph{d}, which according to two small
flares in the Mets\"ahovi data (Harri Ter\"asranta, private communication)
took place around 1995.8 and 1996.7 (just before our first epoch),
respectively.  Note that the apparently curved jet of 0735+178 would appear
more pronounced if components are ejected with different speeds, as seems to
be the case.

  Because of the short time range between our two epochs, the differences in
the positions of the components lie within the errors in the model-fitting,
therefore we can only rely on the observed magnetic field structure, and
previous detections of components motions, to study the possibility of
ballistic motions for the components in 0735+178.

\subsection{Polarization}

  The peak polarization at both epochs and wavelengths is located at the core,
with a degree of polarization between 2 and 4\%, very close to the values
measured by Gabuzda, Wardle, \& Roberts (\cite{De89}) and G94 at 6 and 3.6
cm. However, these values should be regarded as approximate, since maxima in
the images of polarized intensity are not always precisely coincident with
maxima in the total intensity (see also G\'omez et al. \cite{Go98}). Greater
differences between the peaks in total and polarized intensity are found in
the 7 mm map. Some polarized emission is also visible further from the core,
especially at the first epoch at 1.3 cm. In this map the jet between
components \emph{C} and \emph{B} can be traced in polarization, where it has a
rather uniform intensity. For the second epoch at 1.3 cm the core and
component \emph{C} are clearly detected in polarization, and there is also
weaker polarized flux detected at the position of \emph{B}. In the 7 mm
polarization map the core and component \emph{c2} are clearly visible. Another
component appears close to the core, and we have tentatively identified this
component as the polarized counterpart of \emph{d}. Component \emph{c3} is
also observed in polarization, while \emph{c1} is just marginally detected.

  A very similar orientation of the magnetic field vector in the jet is
observed at both epochs in the 1.3 cm maps and in the 7 mm map at November
1996. The magnetic field in the core at our first epoch is oriented toward the
east, in the direction of component \emph{C}. This is consistent with the
values found by Gabuzda, Wardle, \& Roberts (\cite{De89}), although opacity
effects due to the different wavelengths of observation may render this
comparison meaningless.  If the core is optically thick at both wavelengths,
we should consider an extra rotation of 90\( ^{\circ } \) in the polarization
angle when comparing to the remaining optically thin jet. At epoch November
1996, adding the flux of the core and component \emph{d} at 7 mm, we obtain a
very similar value (the same to within the uncertainty) as that measured for
the core at 1.3 cm. The same is true for the second epoch. The core
polarization angle in the 7 mm map shows a slightly different value than at
1.3 cm, while component \emph{d} shows a magnetic field more aligned with the
jet. These differences between the 1.3 cm and 7 mm maps are probably due to
differences in resolution, although they may reflect a change in the opacity
of the core. G94 found a rotation of the polarization angle in the core when
observed at different epochs and frequencies.  These changes in the
polarization angle may be due to ejections of new components with polarization
angles different than the core. Because of the lower resolution, these may
appear in their maps as variations of a single blended component -their core-
similar to what we have experienced when comparing our 1.3 cm and 7 mm maps.

\begin{deluxetable}{lcccccccc}
\tablecolumns{9}
\tablewidth{0pc}
\tablecaption{43 GHz models for 0735+178.\label{table2}}
\tablehead{
&
{\scriptsize \( \alpha  \) }&
{\scriptsize \( \delta  \)}&
\( I \)&
\( p \)&
\( \chi _{m} \)&
\( m \)&
\( r \)&
\( \theta  \)\nl
Component&
(mas)&
(mas)&
(mJy)&
(mJy)&
(\( ^{\circ } \))&
(\%)&
(mas)&
(\( ^{\circ } \))}
\startdata 
\multicolumn{9}{c}{1996.86}\nl
\hline
Core\dotfill&
...&
...&
241&
7&
116&
3&
...&
...\nl
d\dotfill&
0.10&
0.02&
98&
2&
84&
2&
0.10&
80\nl
c3\dotfill&
0.46&
0.11&
26&
4&
29&
15&
0.47&
76\nl
c2\dotfill&
0.87&
0.09&
189&
10&
-6&
5&
0.87&
84\nl
c1\dotfill&
0.96&
0.66&
62&
2&
26&
3&
1.16&
56\nl
b\dotfill&
0.7&
1.5&
48&
...&
...&
...&
1.7&
24\nl
\hline 
\multicolumn{9}{c}{1996.98}\nl
\hline 
Core\dotfill&
...&
...&
370&
...&
...&
...&
...&
...\nl
c\dotfill&
0.78&
0.18&
229&
...&
...&
...&
0.80&
77\nl
b2\dotfill&
0.8&
1.2&
23&
...&
...&
...&
1.4&
34\nl
b1\dotfill&
1.1&
1.7&
23&
...&
...&
...&
2.0&
33\nl
a\dotfill&
3.4&
1.9&
13&
...&
...&
...&
3.9&
61\nl
\enddata
\end{deluxetable}

  The magnetic field structure in the jet between \emph{C} and \emph{B} is
complex, and difficult to interpret unambiguously. One obvious possibility is
that the magnetic field follows the direction of the apparently twisted jet
flow. In this case, the magnetic field throughout this part of the jet would
be longitudinal.  However, another interpretation is possible: that different
features move from the core in different structural position angles, and that
the magnetic field vectors are transverse to the flow direction in \emph{C}
and along the flow direction from the core in \emph{B}. This alternative point
of view is also consistent with the fact the magnetic field in component
\emph{c3} also points back toward the core. In either of these interpretations
the magnetic field in the jet north of \emph{C} is longitudinal.

\section{Theoretical Interpretations and Conclusions}

  Although 0735+178 represents the most abruptly bent jet observed at
milliarcseconds scales to our knowledge, many jets in BL~ Lacs and quasars
appear to be curved. While this is amplified by projection effects, since
relativistic jets in blazars must be pointing close to the line of sight to
explain superluminal motion and related phenomena, it is not clear what causes
the bends of at least 5\( ^{\circ } \) that must be present to account for the
observed twists. 

  One idea (Hardee \cite{Ha87}), which has supporting observational evidence
(Denn \& Mutel \cite{DM98}), is that the jet precesses, with the velocity
vectors of the flow following the bends in the jet axis. Numerical
three-dimensional magnetohydrodynamical simulations (Hardee, Clarke, \& Rosen
\cite{Ha97}) show how Kelvin-Helmholtz (K-H) helical instabilities can develop
from an initial induced precession at the jet inlet, resulting in a twisted
jet with a helical geometry and magnetic field configuration. Such K-H
instabilities can also arise from an external pressure gradient not aligned
with the initial direction of the jet flow.

  Another possibility for jet bending is ballistic precession, such that the
flow velocity is directed radially away from the jet apex, in the manner of
SS~433; this may be suggested by the trajectories of components \emph{K3},
\emph{K4}, \emph{K5}, and \emph{K6} (G94). Alternatively, the nozzle of the
jet could change direction in a more erratic fashion than for
precession. Bends can also occur in response to gradients in the pressure of
the external medium that confines the jet. Another possibility, also suggested
for 0735+178 by G94, is that components only illuminate part of a broad jet
funnel, giving the illusion of a bent trajectory.

  Each possibility for jet bending results in specific predictions. For
example, in the case of erratic change in the nozzle direction, knots should
separate from the core following ballistic trajectories. For ballistic
precession of the nozzle, the different sections of the jet should connect
smoothly and the magnetic field direction should relate to the radius vector
from the core, rather than follow the jet curvature. For precession in which
the flow velocity vector changes direction with the jet axis, moving
components should follow the bends, as should the magnetic field vector. In
addition, the flux density and apparent velocity of the components should
change as the angle between the velocity and the line of sight vary, as in
4C~39.25 (Alberdi et al. \cite{An93}). A precessing jet can be distinguished
from one that changes because of external pressure gradients by the regular
pattern of the bend, which should follow the geometry of a helix in
projection.

  With only 41 days between our two epochs, the expected motions of the
components in the jet of 0735+178 lie within the errors in our model-fitting,
hence we cannot yet conclude whether components on these scales follow
ballistic trajectories or the direction of the jet axis. A ballistic
extrapolation of the position of the components observed in G94 shows that
ballistic motions could explain the observed twisted structure for the
jet. The information provided by the direction of the magnetic field, which
follows the bend near component \emph{B}, suggests that the jet in 0735+178
precesses such that the flow velocity is parallel to the jet axis. Indeed, the
curved jet axis is consistent with the projection of a helix, which would have
to experience an increase in the helical wavelength beyond component \emph{B}
to account for the rather straight jet observed at larger scales (see
G94). However, the interpretation of the magnetic field structure in \emph{B}
and in the jet between \emph{B} and \emph{C} is not entirely clear.  It may be
that the magnetic field in this part of the jet is associated with radial flow
from the core along different position angles, and is longitudinal -directed
back toward the core, suggestive of ballistic motion.

  Further high resolution polarimetric VLBI observations of 0735+178 are
necessary to measure proper motions and determine whether the apparent twisted
structure in this source represents a common curved path for the jet flow or a
superposition of independently ejected ballistically moving components. Such
observations would also make it possible to search for systematic changes in
the jet emission of the sort expected if the nozzle of the jet of 0735+178
precesses.

\begin{acknowledgements}
This research was supported in part by Spain's Direcci\'on General de
Investigaci\'on Cient\'{\i}fica y T\'ecnica (DGICYT), grants PB94-1275 and
PB97-1164, by NATO travel grant SA.5-2-03 (CRG/961228), and by U.S. National
Science Foundation grant AST-9802941.
\end{acknowledgements}


\begin{thebibliography}{1994a}

\bibitem[1993]{An93}Alberdi, A., Marcaide, J. M., Marscher, A. P., Zhang,
Y. F., El\'osegui, P., G\'omez, J. L. \& Shaffer, D. B. 1993, ApJ, 402, 160

\bibitem[1991]{BZ91}B\"a\"ath, L. B. \& Zhang, F. J. 1991, A\&A, 243, 328

\bibitem[1991]{Bal91}B\"a\"ath, L. B., Zhang, F. J. \& Chu, H. S. 1991, A\&A,
250, 50

\bibitem[1974]{carswell 74}Carswell, R. F., Strittmatter, P. A., Williams,
R. D., Kinman, T. D. \& Serkowski, K. 1974, ApJ, 190, L101

\bibitem[1998]{DM98}Denn, G. R. \& Mutel, R. L. 1998, in IAU Colloq. 164:
Radio Emission from Galactic and Extragalactic Compact Sources,
ed. J. A. Zensus, G. B. Taylor, and J. M.  Wrobel, ASP Conference Series,
Vol. 144, 169

\bibitem[1989]{De89}Gabuzda, D. C., Wardle, J. F. C. \& Roberts, D. H. 1989,
ApJ, 338, 743

\bibitem[1994a]{De94}Gabuzda, D. C., Wardle, J. F. C., Roberts, D. H., Aller,
M. F. \& Aller, H.  D. 1994a, ApJ, 435, 128

\bibitem[1994b]{De94b}Gabuzda, D. C., Mullan, C. M., Cawthorne, T. V., Wardle,
J. F. C. \& Roberts, D. H. 1994b, ApJ, 435, 140

\bibitem[1997]{Go97}G\'omez, J. L., Mart\'{\i}, J. M., Marscher, A. P.,
Ib\'a\~nez, J. M. \& Alberdi, A. 1997, ApJ, 482, L33

\bibitem[1998]{Go98}G\'omez, J. L., Marscher, A. P., Alberdi, A., Mart\'{\i},
J. M. \& Ib\'a\~nez, J. M. 1998, ApJ, 499, 221

\bibitem[1987]{Ha87}Hardee, P. E. 1987, ApJ, 318, 78

\bibitem[1997]{Ha97}Hardee, P. E., Clarke, D. A. \& Rosen, A. 1997, ApJ, 485,
533

\bibitem[1998]{Ke98}Kellermann, K. I., Vermeulen, R. C., Zensus, J. A. \&
Cohen, M. H. 1998, AJ, 115, 1295

\bibitem[1995]{Le95}Lepp\"anen, K. J., Zensus, J. A. \& Diamond, P. J. 1995,
AJ, 110, 2479

\bibitem[1997]{sh97}Shepherd, M. C. 1997, in Astronomical Data Analysis
Software and Systems VI, Astron. Soc. Pac. Conf. Proc., 125

\bibitem[1998]{Hi98}Ter\"asranta, H. et al. 1998, A\&AS, 132, 305

\bibitem[1991]{ZB91}Zhang, F. J. \& B\"a\"ath, L. B. 1991, MNRAS, 248, 566

\end{thebibliography}
\end{document}